\documentclass[%
reprint,
amsmath,amssymb,
aps,
]{revtex4-2}
\usepackage{float}
\usepackage{graphicx}
\usepackage{amsmath}
\usepackage{amsfonts}
\usepackage{amssymb}
\usepackage[bookmarks=true,bookmarksopen=true,
bookmarksnumbered=true,
colorlinks=true,linkcolor=black,
citecolor=red]{hyperref}
\usepackage{subfigure}
\usepackage{slashed}
\usepackage{setspace} 
\usepackage{ulem}
\usepackage{braket}

\usepackage{caption}
\usepackage{color}
\allowdisplaybreaks
\usepackage{enumerate}
\usepackage{graphicx}
\usepackage{dcolumn}
\usepackage{bm}

\begin{document}
	
	\preprint{APS/123-QED}

\title{{Prospects for observing a novel photon-induced $H^-H^+$ production  at HL-LHC }}

\author{\small M. A. Arroyo-Ure\~na }
\email{marco.arroyo@fcfm.buap.mx}
\affiliation{\small Facultad de Ciencias F\'isico-Matem\'aticas}
\affiliation{\small Centro Interdisciplinario de Investigaci\'on y Ense\~nanza de la Ciencia (CIIEC), Benem\'erita Universidad Aut\'onoma de Puebla, C.P. 72570, Puebla, Pue., M\'exico,}

\author{\small O. Félix-Beltrán }
\email{olga.felix@correo.buap.mx}
\affiliation{\small Facultad de Ciencias de la Electrónica, Benem\'erita Universidad Aut\'onoma de Puebla, Apartado Postal 1152, C.P. 72570, Puebla, Pue., M\'exico.}

\author{\small J. Hernández-Sánchez }
\email{jaime.hernandez@correo.buap.mx}
\affiliation{\small Facultad de Ciencias de la Electrónica, Benem\'erita Universidad Aut\'onoma de Puebla, Apartado Postal 1152, C.P. 72570, Puebla, Pue., M\'exico.}

\author{\small C. G. Honorato}
\email{carlosg.honorato@correo.buap.mx}
\affiliation{\small Facultad de Ciencias de la Electrónica, Benem\'erita Universidad Aut\'onoma de Puebla, Apartado Postal 1152, C.P. 72570, Puebla, Pue., M\'exico.}

\author{T.A. Valencia-P\'erez}
\email{tvalencia@fisica.unam.mx}
\affiliation{Instituto de F\'isica, 
	Universidad Nacional Aut\'onoma de M\'exico, C.P. 01000, CDMX, M\'exico.}

\begin{abstract}	
    We study the production of charged scalar boson pairs via photon-induced $H^{-}H^{+}$ in $pp$ collisions at the forthcoming High Luminosity Large Hadron Collider (HL-LHC). The predictions are based on the Two-Higgs Doublet Model type-III, since it predicts large branching ratios of the channel decay $H^{\pm}\to \tau^{\pm}\nu_{\tau}$, which is the one we focus on. Motivated by the current excess of events in the process $\mathcal{BR}(t\to H^\pm b)\times \mathcal{BR}(H^\pm\to cb)$ for $M_{H^{\pm}}=130$ GeV, we explore charged Higgs bosons masses $M_{H^{\pm}}$ among 100-150 GeV. According to benchmarks points allowed by experimental constraints, and including a systematic uncertainty of 5$\%$, we obtain for the process of $H^-H^+\to \tau\nu_\tau\tau\nu_\tau\to e\mu+\slashed{E}_T$ a \textit{signal significance} at $5\sigma$ ($3\sigma$) for $M_{H^{\pm}}=100$ GeV ($100\leq M_{H^\pm} \leq 136$ GeV) with 3000 $fb^{-1}$ ($\gtrsim 1000\,fb^{-1}$) integrated luminosity, which would be
the last stage foreseen for Large Hadron Collider machine.
\end{abstract}

\maketitle

\section{Introduction}

According to the laws of classical electrodynamics, two intersecting light beams would not deflect, absorb or disrupt one another. However, effects of quantum electrodynamics allow interactions among photons. 
Four years ago, the ATLAS collaboration reported the observation of photon-induced $W^+W^-$ production ($\gamma\gamma\to W^+W^-$) in proton-proton ($pp$) collisions at $\sqrt{s}=13$ TeV~\cite{ATLAS:2020iwi}. The measurement was performed selecting $e\mu$ pairs that come from the decays $WW\to e\mu \nu_e\nu_\mu$, being in agreement with the theoretical Standard Model (SM) prediction. Inspired by this observation, herein, we suggest a charged scalar particle slightly more massive than the $W^{\pm}$ gauge boson, which can be detected by the same method, being a novel process which could be encouraged at future huntings of the HL-LHC~\cite{Apollinari:2015wtw}. Such a hypothetical particle, denoted as $H^{\pm}$, is  predicted by several SM extensions and, particularly, in the so-called Two-Higgs Doublet Model type III (2HDM-III)~\cite{Atwood:1996vj, Branco:2011iw,Hernandez-Sanchez:2012vxa,Diaz-Cruz:2009ysj,Diaz-Cruz:2004wsi,Arroyo-Urena:2013cyf}, assuming this theoretical framework henceforth. The charged scalar bosons $H^{\pm}$ are associated to new degrees freedom that come from the introduction of a second doublet scalar field to the SM. Interactions of incoming photons in $pp$ collisions for the production of {$H^{-}H^{+}$ pairs ($\gamma\gamma \to H^- H^+$) can provide an invaluable window for the search for such a particle and analyze its features. As well as being irrefutable proof of physics beyond the standard model. In the context of the 2HDM-III, the $\gamma \gamma \to H^- H^+$ interaction proceeds from trilinear and quartic gauge-scalar boson interactions, as shown in Fig.~\ref{interactions}. 
\begin{figure}[!htb]
	\centering
	\includegraphics[width=0.3\textwidth]{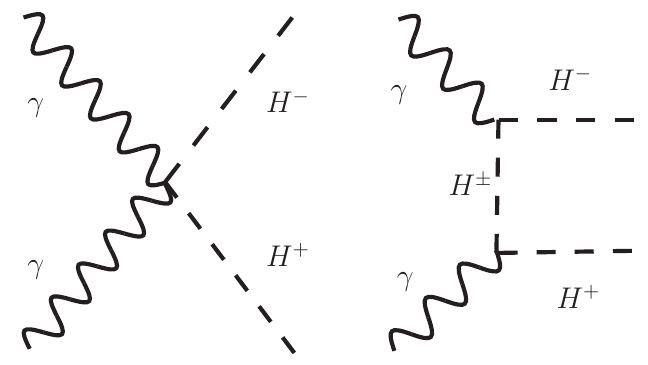}
	\caption{Feynman diagrams for the production of two charged scalar boson via photon fusion in $pp$ collisions.} \label{interactions}	
\end{figure}

In contrast to the production of $\gamma\gamma\to WW\to e\mu+\slashed{E}_T$, the generation of events produced in the process $\gamma\gamma\to H^-H^+\to e\mu+\slashed{E}_T$ is smaller from one to two orders of magnitude. Mainly due to the Yukawa coupling $Y_{H^\pm f_l f_\nu} \propto m_\ell $ which has a significant impact on the cross section $\sigma(\gamma\gamma \to H^- H^+\to e\mu+\slashed{E}_T)$~\cite{Hernandez-Sanchez:2012vxa,Diaz-Cruz:2009ysj,Diaz-Cruz:2004wsi}. Therefore, a larger data set is required to study it copiously. One realistic scenario would certainly be the upcoming HL-LHC. It would be a new stage of the LHC foreseen about 2026 with a center-of-mass energy of 14 TeV. 
Thus, in this work, we are interested in analyzing the prospect for hunting the process $\gamma\gamma\to H^- H^+$ that will produce a final state $e\mu+\slashed{E}_T$, which comes from the decay product of the pair $H^-H^+\to \tau\nu_\tau\tau\nu_\tau\to e\mu+\rm 2 \nu_\tau \nu_e \nu_\mu \to e\mu+\slashed{E}_T$.

In this letter, we show in Sec.~\ref{sec:SecII} the relevant aspects of the 2HDM-III with a particular emphasis on the interactions $H^{\pm}f_i\bar{f}_j$. A brief overview of the model parameter space is also included. In Sec.~\ref{sec:SecIII}, we analyze the charged Higgs boson pairs production via photon fusion, with their subsequent decays into charged leptons plus missing energy transverse (MET) due to neutrinos undetected. Finally, the conclusions are presented in Sec.~\ref{sec:SecIV}.

\section{The model} \label{sec:SecII}
This section presents the relevant aspects of the 2HDM-III. Specifically, we focus on the interactions involved in the main idea of our research, that is, the interactions $H^{\pm}\ell\nu_\ell$, $\gamma\gamma H^- H^+$ and $\gamma H^-H^+$.  


The 2HDM-III includes an additional Higgs doublet regarding to the SM, they can written as $\Phi_a^T=( \phi_{a}^{+},\phi_{a}^{0})$ for $a=1, 2$, with hypercharge $Y=+1$. After the Spontaneous Symmetry Breaking (SSB) they acquire non-zero Vaccum Expectation Values (VEV), and are given as
	\begin{center}
		$\braket{\Phi_a}=\frac{1}{\sqrt{2}}\left(\begin{array}{c}
			0\\
			\upsilon_a
		\end{array}\right),\,\,a=1,\,2.$
		\end{center}
According to our previously and exhaustive analysis of the model, only the auto-interactions of Higgs bosons and Yukawa couplings are modified~\cite{Hernandez-Sanchez:2012vxa,Diaz-Cruz:2009ysj}. Therefore, the interactions $\gamma\gamma H^- H^+$ and $\gamma H^-H^+$ are unchanged and well known~\cite{Branco:2011iw}.
The relevant parameters of the Higgs potential of the model are the mixing angle $\alpha$ related to the mass matrix of CP-even sector, the angle $\beta$ which defines the ratio of the two VEVs  ($\tan\beta=\upsilon_2/\upsilon_1$), and the physical masses of the scalar neutral, pseudoscalar and charged Higgses, $M_h$, $M_H$, $M_A$ and $M_{H^{\pm}}$, respectively.
Thus, the relevant interactions to study are given by the 2HDM-III Yukawa Lagrangian~\cite{Diaz-Cruz:2004wsi,Hernandez-Sanchez:2012vxa}:
\begin{eqnarray}
	\label{YukawaLagrangian} 
	-\mathcal{L}_Y &=& Y_{1}^{u} \bar{Q}_{L} \tilde{\Phi}_{1} u_{R} + Y_{2}^{u} \bar{Q}_{L} \tilde{\Phi}_{2} u_{R} + Y_{1}^{d} \bar{Q}_{L} \Phi_{1} d_{R}\\ \nonumber
	&+& Y_{2}^{d} \bar{Q}_{L} \Phi_{2} d_{R} + Y_{1}^{l} \bar{L}_{L} \tilde{\Phi}_{1} l_{R} + Y_{2}^{l} \bar{L}_{L} \tilde{\Phi}_{2} l_{R},
\end{eqnarray}
where  $\tilde{\Phi}_{a} = i\sigma_2 \Phi_{a}^{*}$ $(a=1, 2)$.  After SSB, the fermion mass matrices are defined as: $M_f = \frac{1}{\sqrt{2}} \left( v_1 Y_{1}^{f} + v_2 Y_{2}^{f} \right),\; f=u,d,\ell$.	
We assume both Yukawa matrices have a four-zero texture form and are Hermitian. In order to obtain the diagonal mass matrices $\bar{M}_f$, we apply the following bi-unitary transformation $\bar{M}_f = V_{fL}^{\dagger} M_f V_{fR}= \tfrac{1}{\sqrt{2}} \left( v_1 \tilde{Y}_{1}^{f} + v_2 \tilde{Y}_{2}^{f} \right), \,
\tilde{Y}_{a}^{f} = V_{fL}^{\dagger} Y_{a}^{f} V_{fR}$~\cite{Felix-Beltran:2013tra}. Once the diagonalization is done, following our previous analysis~\cite{Hernandez-Sanchez:2012vxa,Diaz-Cruz:2009ysj}, we obtain:  
\begin{equation}\label{RotateYukawas}
	\left[ \tilde{Y}_a^f \right]_{ij} = \frac{\sqrt{m_i m_j}}{v} \left[ \chi_a^f \right]_{ij},
\end{equation}
where the parameters $\chi$'s are  dimensionless quantities of the present model. 
 As a consequence, we have non-zero off-diagonal elements and hence flavor-changing neutral interactions at the tree level. 
The relevant interactions we are interested in, come from Eqs.~\eqref{YukawaLagrangian}-\eqref{RotateYukawas} and are given by:
\begin{eqnarray}
	-\mathcal{L}^{H^{\pm}f_\ell\bar{f}_\nu}&=&  \frac{\sqrt{2} m_{\ell_j}}{v} \mathcal{Z}_{ij}^\ell \bar{\nu}_L l_R H^{+} + \textrm{H.c.},
\end{eqnarray}
where 
\begin{eqnarray}
	\mathcal{Z}_{ij}^{\ell} &=& \left[ \mathcal{Z} \frac{m_{\ell_i}}{m_{\ell_j}} \delta_{ij} -\frac{F(\mathcal{Z})}{\sqrt{2}} \sqrt{\frac{m_{\ell_i}}{m_{\ell_j}}} \chi_{ij}^{\ell} \right],
\end{eqnarray}
with $F(\mathcal{Z})=\sqrt{1+\mathcal{Z}^2}$,  and $\mathcal{Z}$ definitions given in Table~\ref{XYZ} for the four kinds of versions. In the limit $\chi_{ij}^\ell$ $\to$ 0, we recover the traditional four types  of 2HDM. 
 \begin{table}[htp]
	\begin{center}
		\begin{tabular}{|c|c|}
			\hline
			2HDM-III & $\mathcal{Z}$\\
			\hline \hline
			Type I-like &   $- \cot \beta$ \\
			\hline
			Type II-like &  $ \tan \beta$ \\
			\hline
			Lepton Specific-like& $ \tan \beta$ \\
			\hline
			Flipped-like   & $ -\cot \beta$ \\
			\hline
		\end{tabular}
	\end{center}
	\caption{Definitions of the $\mathcal{Z}$ parameter for the different model types.}
	\label{XYZ}
\end{table}
%

\subsection*{Constraints on the 2HDM parameter space \label{se:conts}}
From now on we focus on the particular case of the 2HDM-III type II-like (simply called as throughout the text: 2HDM-III). For this version, a comprehensive analysis of the parameter space is reported in Ref.~\cite{Arroyo-Urena:2024soo}, which was done considering several physical observables, namely: $i)$ LHC Higgs boson data (and its HL-LHC projection) and $ii)$ Lepton Flavor-Violating processes (LFVp).
The parameters that have a direct impact on our predictions are: $\cos(\alpha-\beta)$, $\tan\beta$ and $\chi_{33}^{\ell}\equiv\chi_{\tau\nu}$. 
According to Ref.~\cite{Arroyo-Urena:2024soo}, the allowed region in the $\cos(\alpha-\beta)-\tan\beta$ plane is presented in Fig.~\ref{fig:parameter_space}.
\begin{figure}[!htb]
	\centering
	{\includegraphics[scale=0.25]{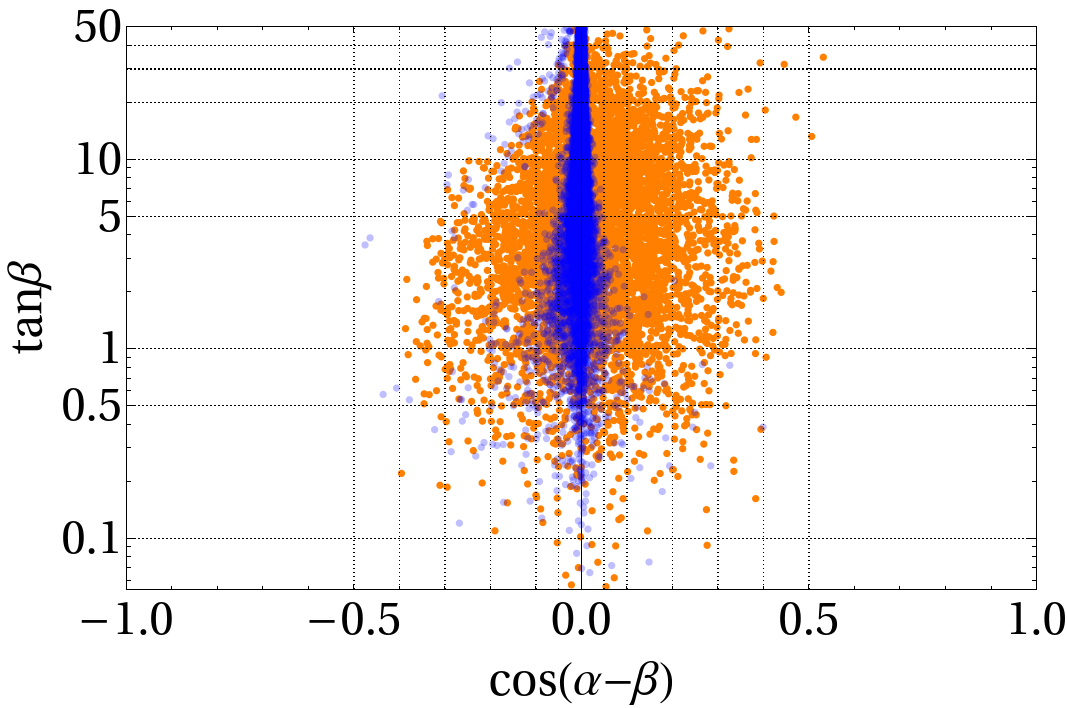}}	
	\caption{Parameter space in the $\cos(\alpha-\beta)-\tan(\beta)$ plane. Blue (orange) points represent the allowed by LHC Higgs boson data (LFVp). Figure taken from~\cite{Arroyo-Urena:2024soo}.}
\label{fig:parameter_space}
\end{figure}
Thus, we define three benchmark points (BP) to be considered in the predictions presented in Sec.~\ref{sec:SecIII}, namely,
\begin{enumerate}[A.]
        \item BP1: $\tan\beta=10$, $\cos(\alpha-\beta)=-0.2$,
	\item BP2: $\tan\beta=20$, $\cos(\alpha-\beta)=-0.1$,
	\item BP3: $\tan\beta=2$, $\cos(\alpha-\beta)=0.1$.
\end{enumerate}
In all the BPs, we set $\chi_{\tau\nu}=1$~\cite{Hernandez-Sanchez:2021fgj,Hernandez-Sanchez:2020vax}.

\section{Collider analysis \label{sec:SecIII}}
To explore the prospects for detecting the process $\gamma\gamma\to H^- H^+$, we use a Monte Carlo generator to obtain a sample of simulated events. We also present the strategy to separate the signal from the background.
Let us first present the signal and the main SM background processes as follows:
\begin{itemize}
	\item Signal: We are interested in searching for a final state $e\mu+\slashed{E}_T$ coming from the decay of two charged scalar boson produced through photon fusion in $pp$ collisions at the HL-LHC. Feynman diagrams that contribute in the production of a $H^-H^+$ pair with their subsequent decays are shown in Fig.~\ref{FeynmanDiagramsContribution}. 
	\item Background: The main SM background processes come from quark-induced $WW$ production ($qq\to W^-W^+$), the electromagnetic reaction $\gamma\gamma\to W^+ W^-$ and from other photon-induced processes. Mainly di-lepton production $\gamma\gamma\to \ell\ell$ which are reduced by selecting different flavor lepton pairs, keeping a smaller contribution from $\gamma\gamma\to \tau^-\tau^+$ production with lepton $\tau$ decays.
\end{itemize}
\begin{widetext}
\begin{figure}[!htb]
	\centering
	\includegraphics[width=1\textwidth]{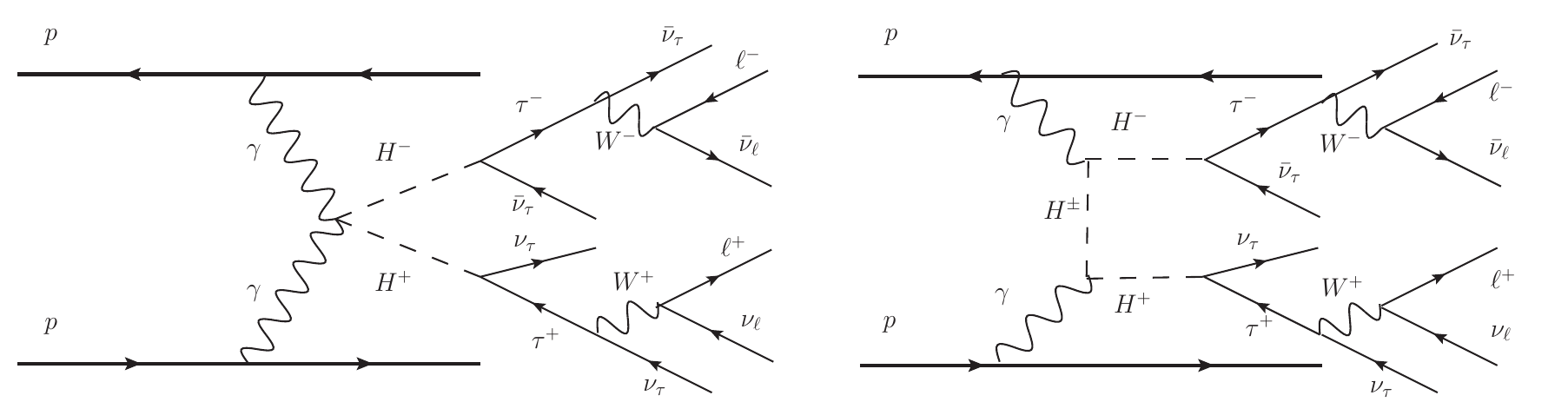}
	\caption{Feynman diagrams for the production of two charged scalar boson via photon fusion in $pp$ collisions.} \label{FeynmanDiagramsContribution}	
\end{figure}
\end{widetext}
Motivated by the relatively recent report by ATLAS collaboration~\cite{ATLAS:2023bzb} on an excess of events at $3\sigma$ in the process $\mathcal{BR}(t\to H^\pm b)\times \mathcal{BR}(H^\pm\to cb)$ for $M_{H^{\pm}}=130$ GeV, we focus on the mass range $M_{H^{\pm}}\in[100,\,150]$ GeV. In Ref.~\cite{Arroyo-Urena:2024soo} the region of the parameter space that accommodates such an excess is studied. Thus, in this work we assume that the 2HDM-III satisfies this experimental constraint. 
In Fig.~\ref{XSallModels} the production cross section~\footnote{Concerning our computation scheme, we first implement the model via \texttt{FeynRules}~\cite{Alloul:2013bka} to obtain the UFO files~\cite{Degrande:2011ua} for \texttt{MadGraph5}~\cite{Alwall:2011uj} which is interfaced with \texttt{Pythia8}~\cite{Bierlich:2022pfr} for parton showering and \texttt{Delphes}~\cite{deFavereau:2013fsa} for detector response. The default HL-LHC card~\cite{HL-LHC_card} and \texttt{NN23NLO}~\cite{Ball:2013hta} PDF set were used in the simulations.} of the signal as a function of the charged scalar boson mass $M_{H^{\pm}}$ is shown. We consider a center-of-mass energy of the $10\%$ (700 GeV by photon)~\cite{Bertulani:2005ru}, which LHC should reach.    
\begin{figure}[!htb]
	\centering
	{\includegraphics[scale=0.22]{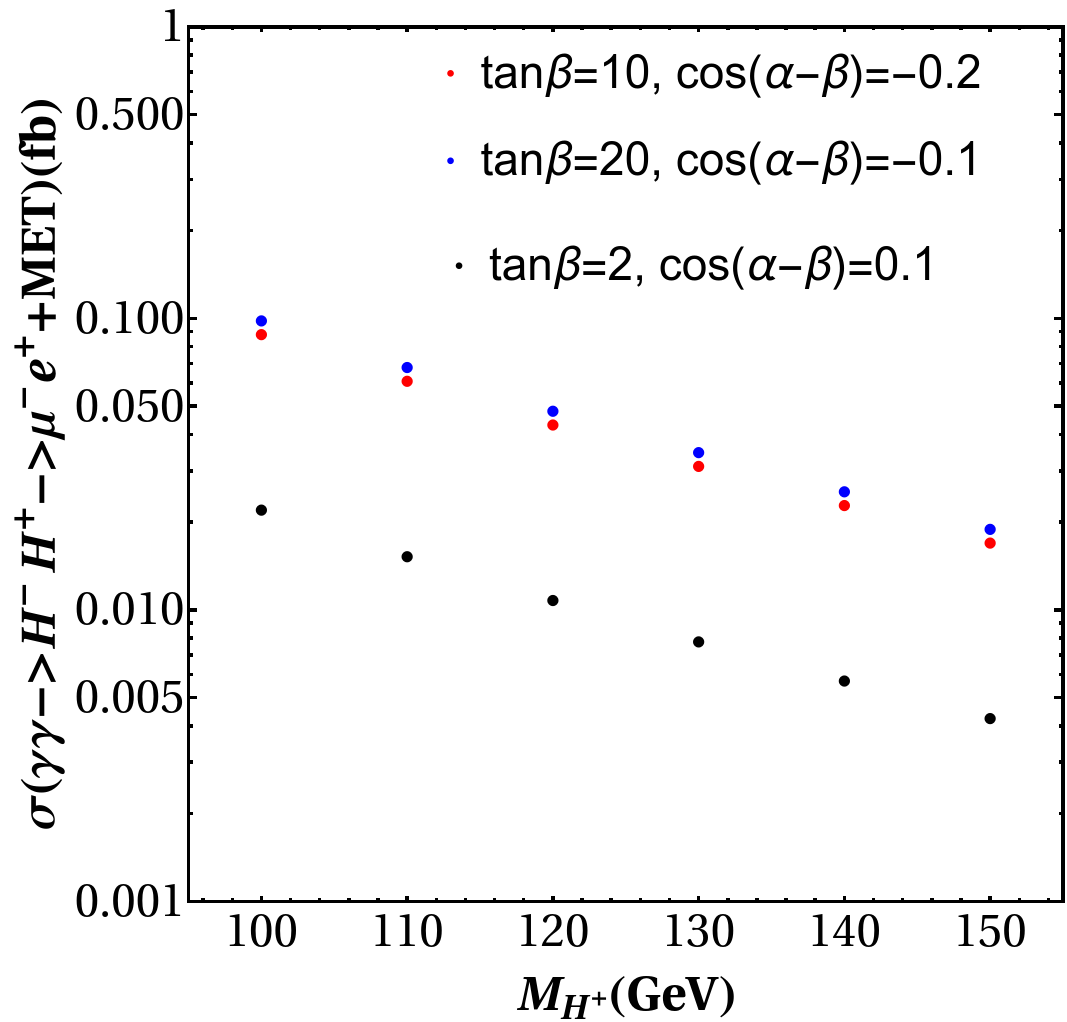}}
		\caption{Production cross section of the signal $\sigma(\gamma\gamma\to H^- H^+\to \tau^+ \nu_{\tau}\tau^- \bar{\nu}_{\tau}\to e^+\mu^-++\rm E_T \hspace{-.4 cm} / \hspace{.2 cm})$ as a function of the charged scalar boson mass $M_{H^{\pm}}$. Red (blue, black) points correspond to BP1 (BP2, BP3).}\label{XSallModels}
\end{figure}

Considering an integrated luminosity $\mathcal{L}_{\rm int}=300$ fb$^{-1}$, in the best case (BP2), the number of events produced for $M_{H^\pm}=100$ GeV is $N_S=30$, which seems discouraging. However, for the HL-LHC the situation is a bit more promising, reaching a number of events in the interval 60-300, enough to claim a possible discovery of the signal proposed, as will be seen later.
In order to isolate the signal from the SM background, we performed a Multivariate Analysis using \textit{Boosted Decision Trees}~\cite{Hastie:2009itz,doi:10.1142/9789811234033_0002} (BDT). The BDT training is computed using Monte Carlo simulated data. The signal and background processes are scaled to the expected number of events, which is calculated through the cross sections and the integrated luminosity. The BDT selection is optimized for each $M_{H^\pm}\in [100,\,150]$ GeV to maximize the \textit{signal significance}, defined as $\mathcal{S}=N_S/\sqrt{N_S+N_B+(\kappa\cdot N_B)^2}$~\cite{Cowan2021}, where $N_S$ and $N_B$ are the number of signal and background candidates, respectively. The factor $\kappa$ stands for a realistic $5\%$ systematic uncertainty. The main hyperparameters for the training are: $i)$ Number of trees \texttt{NTree=50}, $ii)$ maximum depth of the decision tree \texttt{MaxDepth=5}, and $iii)$ maximum number of leaves \texttt{MaxLeaves=5}; the remaining parameters are set to their default values.  We also compute a Kolmogorov-Smirnov test (KS). We observe that the KS test value shows signs of no over-training: 0.12 for the background and 0.15 for the signal. The BDT classifiers were trained using variables related to the kinematic of the final state particles, including the transverse momentum $p_T$, pseudo-rapidity $\eta$, transverse masses of charged leptons $M_T(\ell)$ $(\ell=e,\,\mu)$, angular distance between the muon and electron $\Delta R$, missing energy transverse $\slashed{E}_T$.
Despite having a signal that is difficult to isolate, we are able to highlight it against the background, which is manifested in the discriminant shown in Fig.~\ref{fig:DiscriminantBDT}. In this analysis, we use the BP2 and $M_{H^{\pm}}=100$ GeV. It is important to mention that we also performed a cut-and-count baseline analysis via \texttt{MadAnalysis}~\cite{Conte:2012fm} to compare with BDT. However, this method was not effective.  Fig.~\ref{fig:distributions} shows the most discriminating characteristics according to BDT. In contrast, it is evident that through ``straight" cuts, a significant number of signal events is removed, which considerably reduces the \textit{signal significance}.
\begin{figure}[!htb]
	\centering
	\includegraphics[width=0.42\textwidth]{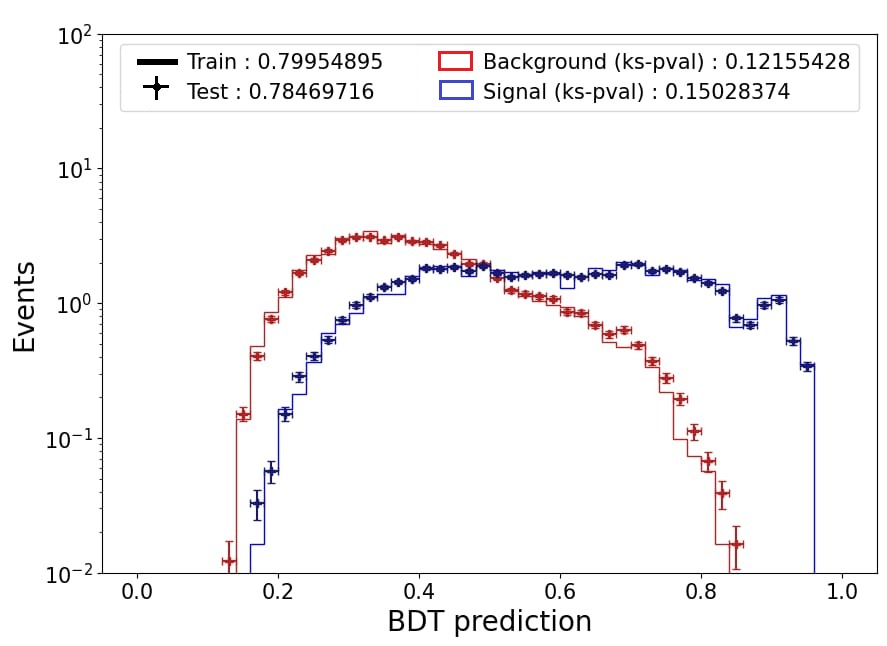}
	\caption{Discriminant for signal and background data generated by BDT. We consider the BP2 and $M_{H^+}=100$ GeV.} \label{fig:DiscriminantBDT}	
\end{figure}
\begin{figure}[!htb]
	\centering
	\includegraphics[scale=0.13]{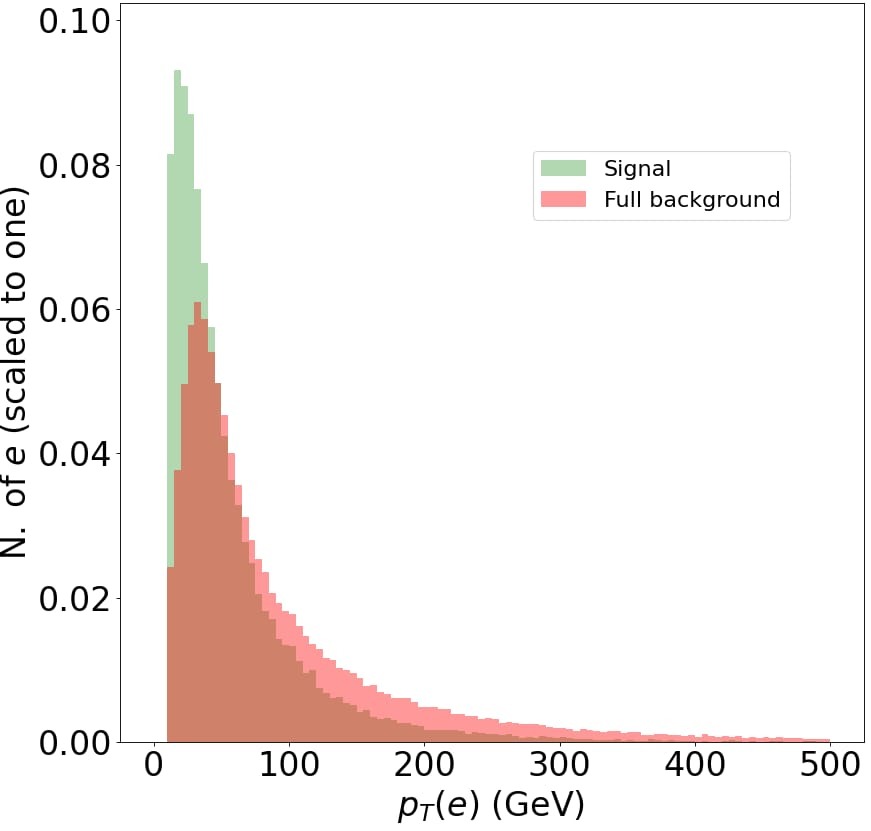}
	\includegraphics[scale=0.13]{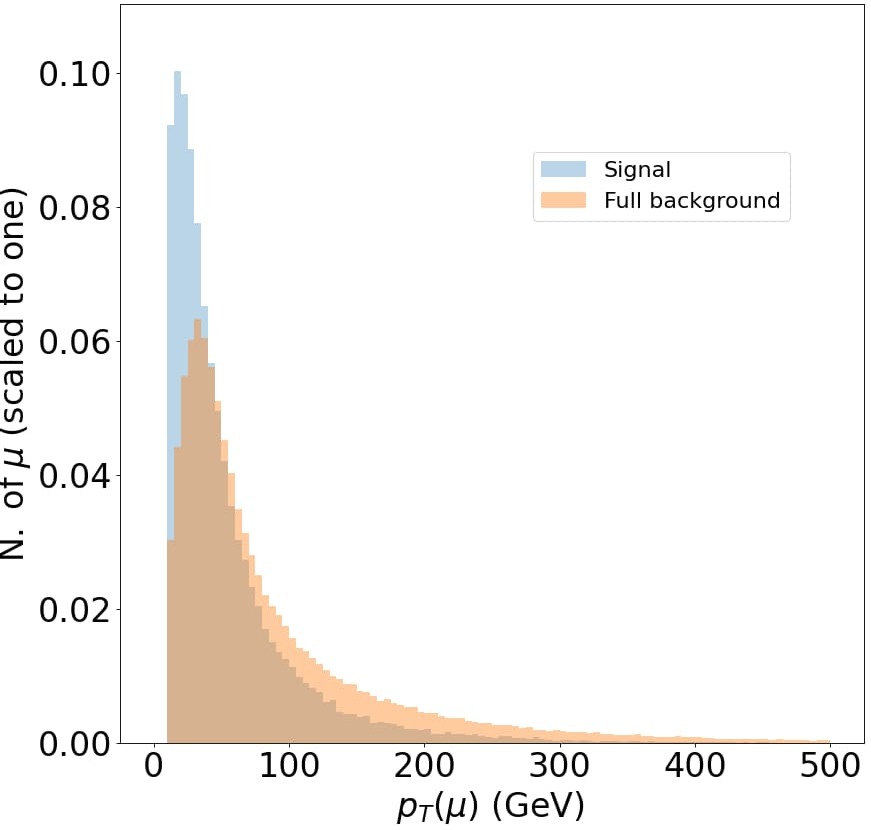}
	\includegraphics[scale=0.26]{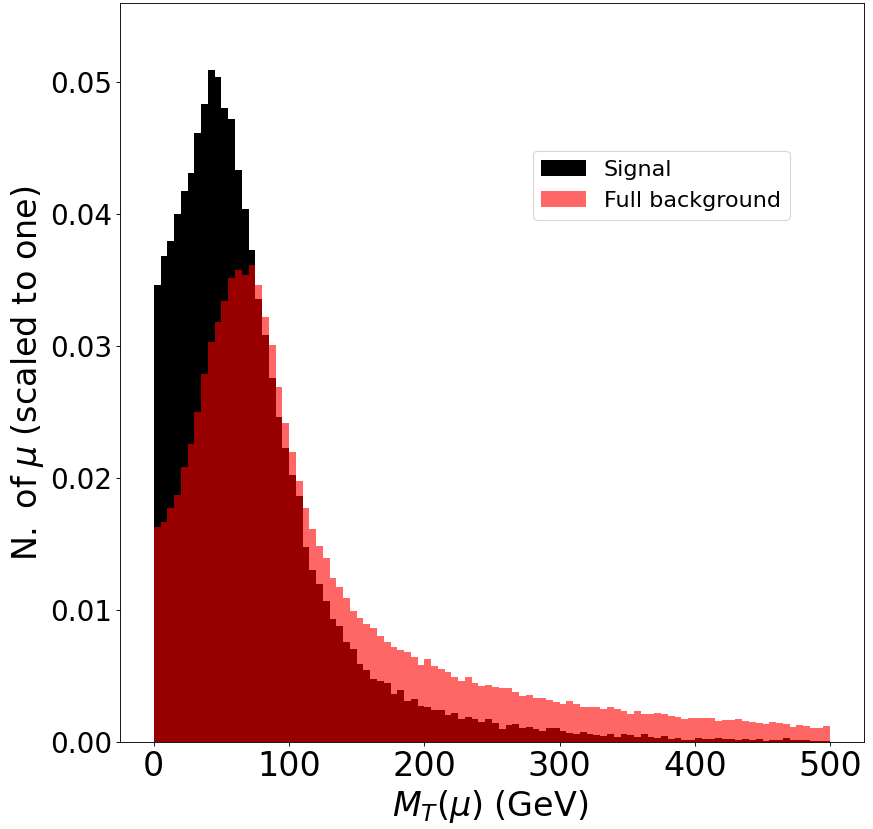}
	\includegraphics[scale=0.13]{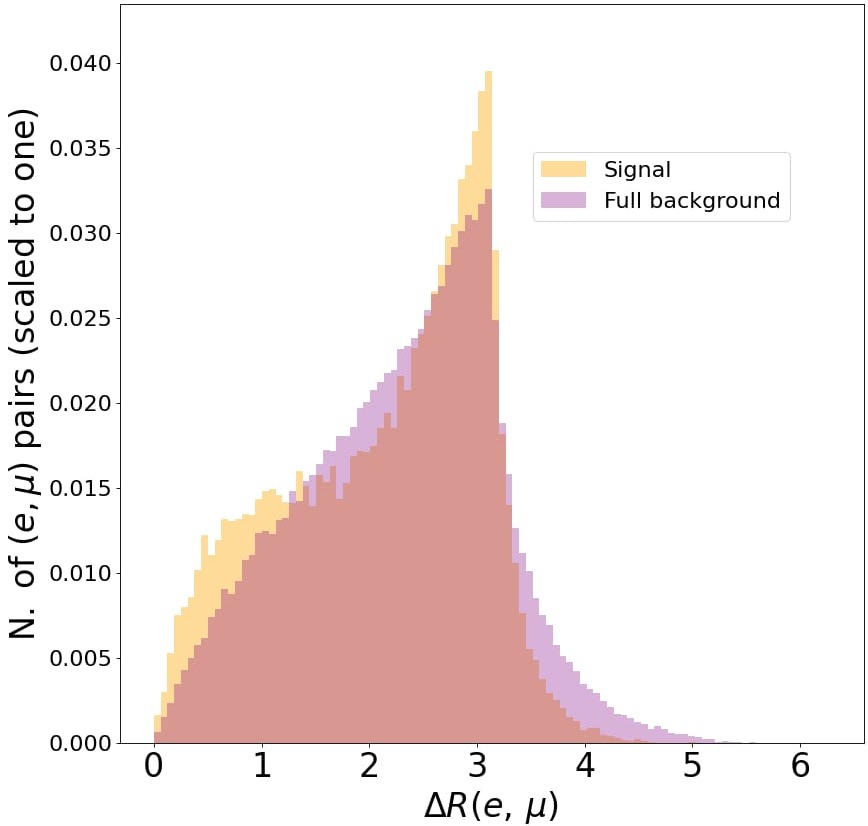}
	\caption{Normalized distributions: (a) transverse momentum of the electron, (b) transverse momentum of the muon, (c) transverse mass of the muon: $M_T(\mu)=\sqrt{2p_\mu^T\slashed{E}_T(1-\cos\phi_{\mu\slashed{E}_T})}$, and (d) angular distance $\Delta R$ between the electron and muon.}\label{fig:distributions}
\end{figure}
Based on the previous analysis, we show in Fig.~\ref{fig:significance} our most important results, \textit{i.e.}, the \textit{signal significance} $\mathcal{S}$, which is presented as a function of the integrated luminosity and the charged Higgs boson mass $M_{H^{\pm}}$. Fig.~\ref{fig:significance}(a) corresponds to BP1 while Fig.~\ref{fig:significance}(b) represents BP2.
\begin{figure}[!htb]
	\centering
	\subfigure[]{\includegraphics[width=0.5\textwidth]{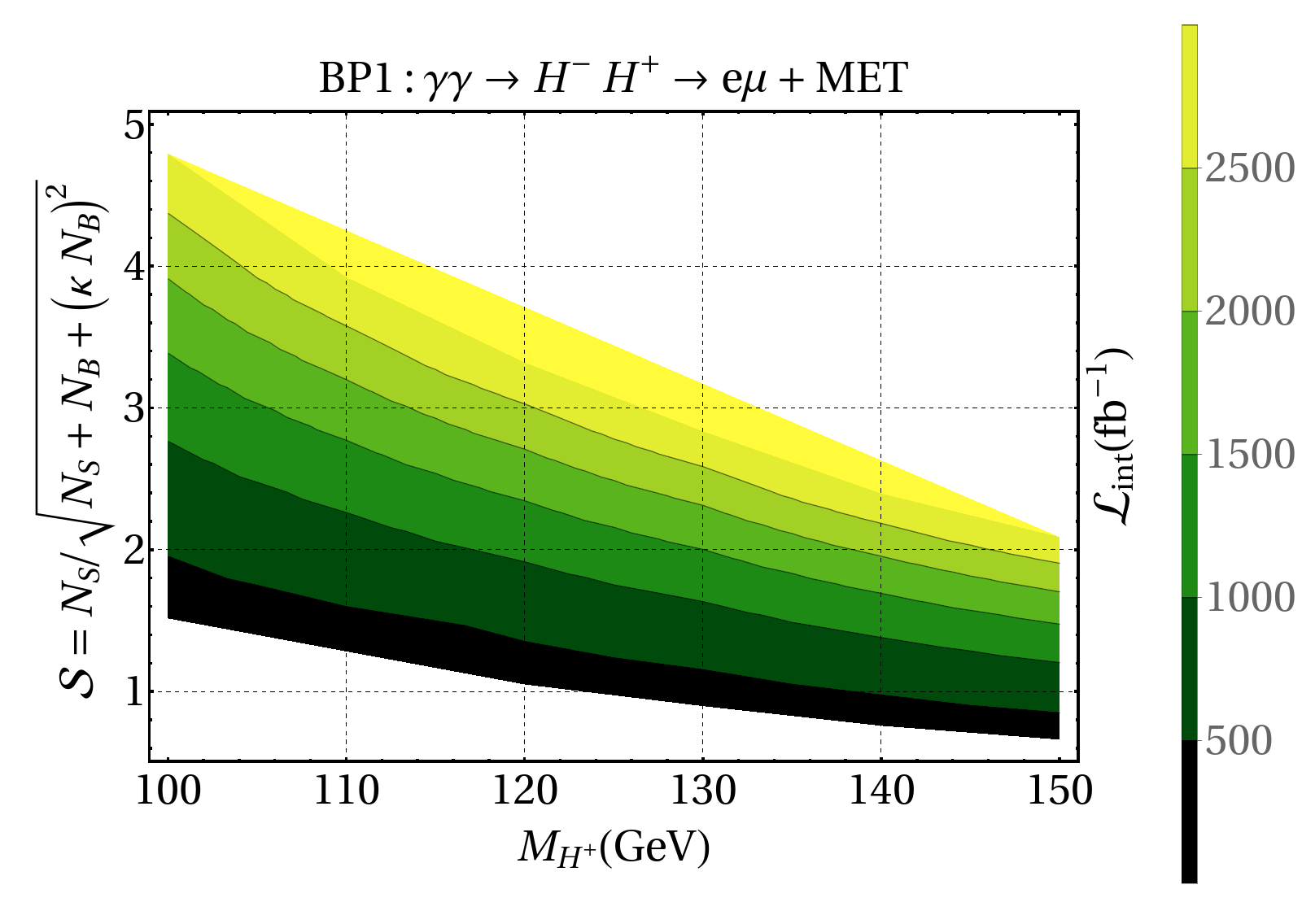}}
		\subfigure[]{\includegraphics[width=0.5\textwidth]{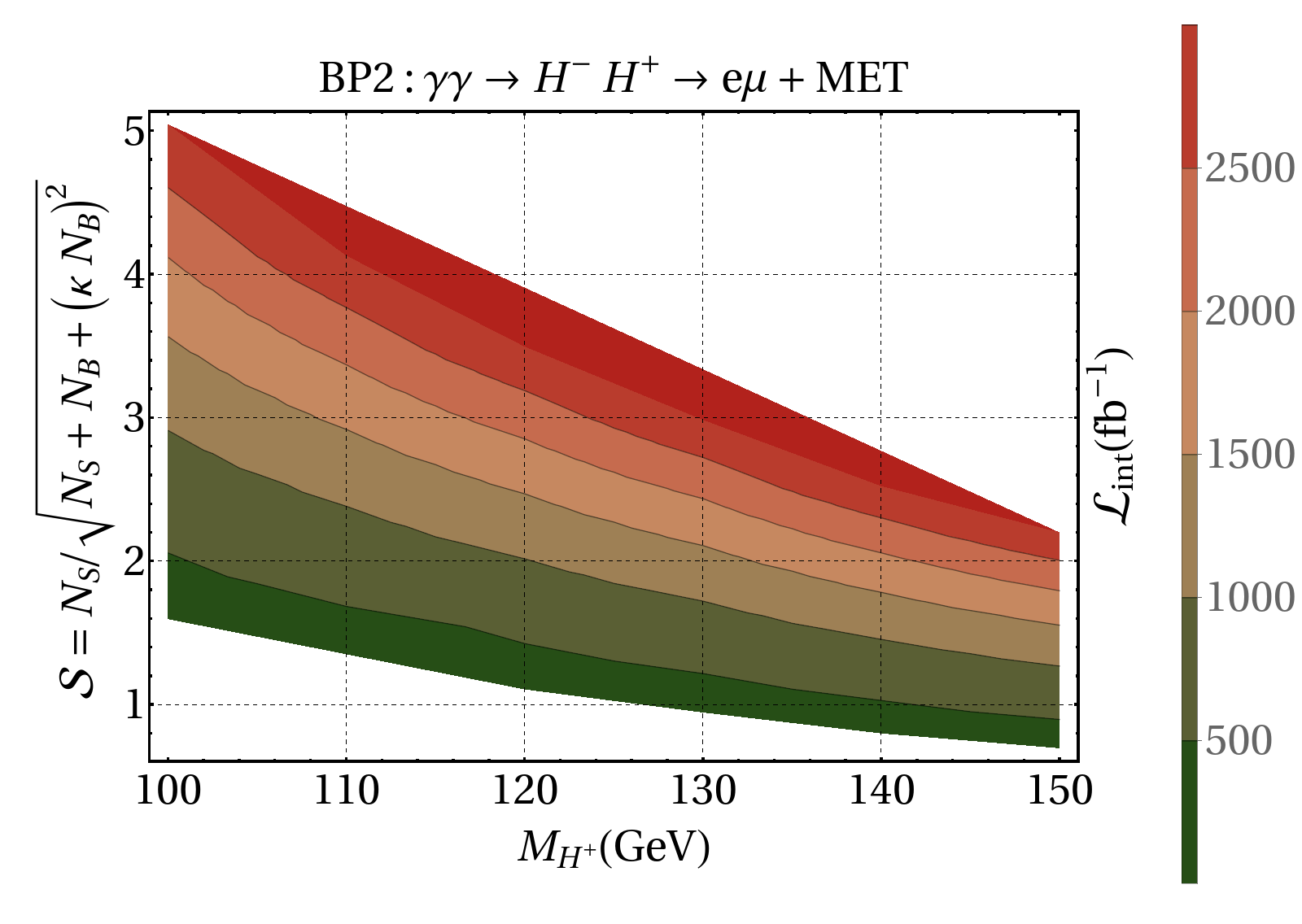}}
	\caption{Signal significance as a function of the integrated luminosity $\mathcal{L}_{\rm int}$ and the charged Higgs boson mass $M_{H^{\pm}}$. (a) BP1 and (b) BP2.} \label{fig:significance}	
\end{figure}
From Fig.~\ref{fig:significance}(b), we observe a possible discovery of a pair $H^-H^+$ for $M_{H^{\pm}}=100$ GeV and $\mathcal{L}_{\rm int}\approx 3000$ fb$^{-1}$. Meanwhile, a potential evidence at the level of $3\sigma$ is found for the interval of masses $M_{H^{\pm}}\in 100-136$ GeV and $\mathcal{L}_{\rm int}\gtrsim 1000$ fb$^{-1}$. For BP1 we obtain very similar results as for BP2, the difference is close to $5\%$. With the HL-LHC era, our results represent an accessible opportunity to observe a charged scalar in a relatively short time.It should be noted that for BP3 a \textit{statistical significance} of $1.7\sigma$ is obtained for $M_{H^{\pm}}=100$ GeV, which encourages searching for this scenario in the High-Energy LHC project~\cite{Benedikt:2018ofy}. 
\section{Conclusions}\label{sec:SecIV}
We have presented an analysis of the photon-induced production of $H^-H^+$ pairs ($\gamma\gamma \to H^-H^+$) and the prospects for their detection in $pp$ collisions at the future HL-LHC. This is a great scenario because of the large amount of data it will collect, compared to that obtained by the LHC. Our results are based on the theoretical framework of the Two-Higgs Doublet Model type-III, which predicts the existence of charged scalar bosons $H^{\pm}$, our object of study.  The final state on which we focus is $e\mu$ plus missing energy transverse due to undetected neutrinos, whose source is the decays of the $H^-H^+$ pairs. The main motivation to explore this signature was the detection of pairs $W^-W^+$ by the ATLAS collaboration, through the method also used in this work. 
Considering three benchmark points (allowed by experimental constraints), we found promising results that can be brought to experimental scrutiny. In particular, we predict a possible discovery of $H^-H^+$ pairs at $5\sigma$, for $M_{H^{\pm}}=100$ GeV and $\mathcal{L}_{\rm int}\approx 3000$ fb$^{-1}$ by considering BP2: $\tan\beta=20$, $\cos(\alpha-\beta)=-0.1$, and $\chi_{\tau\nu}=1$. Similar results were found for BP1: $\tan\beta=10$, $\cos(\alpha-\beta)=-0.2$, and $\chi_{\tau\nu}=1$. Taking into account BP2 (BP1) and integrated luminosities $\mathcal{L}_{\rm int}\gtrsim 1000$ $fb^{-1}$, we also predict \textit{significances} of $3\sigma$ for the mass range 100-136 GeV (100-133 GeV). The least favored situation occurs for BP3, which requires integrated luminosities higher than expected at the HL-LHC. For this scenario, we obtain $1.7\sigma$ for $M_{H^{\pm}}=100$ GeV. However, if the systematic uncertainties ($\kappa=5\%$ considered in this work) were improved, the \textit{signal significance} would increase slightly for all BP's analyzed in this research, opening the possibility of testing BP3. 
\subsection*{Acknowledgments}
The work of Marco A. Arroyo-Ure\~na and T. Valencia-P\'erez is supported by ``Estancias Posdoctorales por M\'exico (SECIHTI)'' and ``Sistema Nacional de Investigadores e Investigadoras'' (SNII-SECIHTI). T.V.P. acknowledges support from the UNAM project PAPIIT IN111224 and the SECIHTI project CBF2023-2024-548. OF-B, CGH and JH-S thank the support of SNII-SECIHTI, VIEP-BUAP and PRODEP, México.
\bibliography{biblio}
\end{document}